\documentclass[
 pra,
 amsmath,amssymb,
 aps,
 twocolumn,
 notitlepage,
 superscriptaddress
]{revtex4-2}

\usepackage{float}
\makeatletter
\let\newfloat\newfloat@ltx
\makeatother

\newcommand{\eSWAP}{{\rm eSWAP}}
\newcommand{\qsharp}{Q\#}
\usepackage[colorlinks=true]{hyperref}

\usepackage[caption=false]{subfig}

\usepackage{styles}
\usepackage{fancyhdr}
\usepackage{txfonts}
\usepackage{xurl} 

\usepackage{tikz}
    \usetikzlibrary{trees}
    
\usepackage{xcolor}
    \definecolor{cud-bluish-green}{RGB}{0,158,115}
    \colorlet{comment}{cud-bluish-green!30!black!70}
    
\usepackage{algpseudocode}
    \newcommand{\linecomment}[1]{\State \(\triangleright\) {\footnotesize #1} \normalsize}    \newcommand{\seccomment}[1]{%
        \vskip0.5em%
        \State {\color{comment} \footnotesize \(\blacksquare\) \emph{#1}} \normalsize
    }

\usepackage{listings}
\usepackage{qsharp}

\begin{document}

\title{Advancing Hybrid Quantum–Classical Computation with Real-Time Execution}

\author{Thomas Lubinski}
\affiliation{Quantum Circuits Inc, 25 Science Park, New Haven, CT 06511, USA}

\author{Cassandra Granade}
\affiliation{Microsoft Corporation, Quantum Architectures and Computation Group, Redmond, WA 98052, USA}

\author{Amos Anderson}
\affiliation{Quantum Circuits Inc, 25 Science Park, New Haven, CT 06511, USA}

\author{Alan Geller}
\affiliation{Microsoft Corporation, Quantum Architectures and Computation Group, Redmond, WA 98052, USA}

\author{Martin Roetteler}
\affiliation{Microsoft Corporation, Quantum Architectures and Computation Group, Redmond, WA 98052, USA}

\author{Andrei Petrenko}
\affiliation{Quantum Circuits Inc, 25 Science Park, New Haven, CT 06511, USA}

\author{Bettina Heim}
\affiliation{Microsoft Corporation, Quantum Architectures and Computation Group, Redmond, WA 98052, USA}


\date{\rule[15pt]{0pt}{0pt}\today}
             
\begin{abstract}

\vspace{0.0cm}
The use of mid-circuit measurement and qubit reset within quantum programs has been introduced recently and several applications demonstrated that perform conditional branching based on these measurements.  In this work, we go a step further and describe a next-generation implementation of classical computation embedded within quantum programs that enables the real-time calculation and adjustment of program variables based on the mid-circuit state of measured qubits. A full-featured Quantum Intermediate Representation (QIR) model is used to describe the quantum circuit including its embedded classical computation. This integrated approach eliminates the need to evaluate and store a potentially prohibitive volume of classical data within the quantum program in order to explore multiple solution paths. It enables a new type of quantum algorithm that requires fewer round-trips between an external classical driver program and the execution of the quantum program, significantly reducing computational latency, as much of the classical computation can be performed during the coherence time of quantum program execution. We review practical challenges to implementing this approach along with developments underway to address these challenges. An implementation of this novel and powerful quantum programming pattern, a random walk phase estimation algorithm, is demonstrated on a physical quantum computer with an analysis of its benefits and feasibility as compared to existing quantum computing methods.
\end{abstract}


\maketitle

\tableofcontents
    

\pagestyle{fancy}

\renewcommand{\headrulewidth}{0.0pt}
\lhead{}
\rhead{\thepage}

\renewcommand{\footrulewidth}{0.4pt}
\cfoot{}

\lfoot{Advancing Hybrid Quantum–Classical Computation with Real-Time Execution}

\rfoot{\today}
\vspace{0.5cm}


\section{Introduction}
\label{sec:introduction}

Over the past decade, quantum computers have become more advanced and  accessible to users. Quantum applications
are theoretically capable of addressing a limited set of computational challenges in an exponentially accelerated time frame \cite{Arute2019-mk}.
Quantum computing research has resulted in hundreds of algorithms shown to function on near-term quantum computing systems \cite{preskill2018quantum,jordan_zoo,bharti2021noisy}.
Recent work has shown that these algorithms offer the potential for quantum advantage over classical computing in specific domains \cite{Arute2019-mk, Zhong2020-rk}.

Progress towards quantum advantage has been hindered, however, by significant challenges in the noisy intermediate-scale quantum computing regime (\emph{a.k.a.} NISQ) \cite{preskill2018quantum,Ferracin2019-ou}. Current generation gate-model devices are restricted to a small number of qubits, and can only execute a limited number of instructions before ``noise'' or gate error dominates.
To address these limitations, creative algorithms such as VQE and QAOA have been developed that take advantage of quantum and classical devices working in tandem.
In these hybrid approaches, a classical computer program iteratively invokes a quantum processor to execute a small part of the algorithm (with some exponential speedup). Bits of problem data are passed to the quantum processor and results returned to the classical processor, which makes decisions about the next batch of quantum instructions to execute and assembles a solution from parts.

However, there is a fundamental limit to how well this approach to hybridizing quantum and classical computing can scale.
Treating quantum and classical processors as disjoint physical instruments, each with their own data transfer pipeline and computational interface, constrains hybrid algorithms to repeatedly switching between contexts and exchanging intermediate data between devices.
This high latency means classical decisions can not be made to influence evolution of quantum state before qubits decohere.

\vspace{0.2cm}

In this paper, we describe a new class of hybrid program in which elements of classical computation are embedded directly within the quantum program and execute in the same time domain as the quantum operations. This approach delivers a compelling advantage, reducing latency of data exchange by orders of magnitude and providing flexibility in controlling the quantum state during execution. We delineate the characteristics of this new type of hybrid quantum/classical computation, the hardware and software required to enable it, and the opportunities it affords. 

This capability requires a quantum computer that is able to execute a series of quantum operations commingled with some (classical) computation that uses the results of previous operations to compute new results that may affect the next iteration or series of quantum operations.
The most important aspect of this implementation is that the classical computation is performed \emph{without terminating execution of the quantum program and discarding the qubit state} or returning to the classical computer for those computations.
Many small classically driven adjustments to the quantum state can be made based on measurements performed in the middle of the program, resulting in a program that is adaptive in nature.

Unlike a classical computer with its programming constructs such as variables, arithmetic computation and looping, a quantum computer is typically implemented using highly specialized hardware and firmware optimized to generate complex sequences of high-resolution microwave or laser pulses to manipulate sensitive and fragile quantum states.
These control systems have only a limited ability to perform integrated classical computation under tight time constraints \cite{corcoles2021dyncircuits,ella_2022}.
We will discuss enhancements that may be necessary to fully enable this new form of hybrid program.

This paper demonstrates a first step towards fully general, tightly integrated quantum/classical processing, enabling new types of quantum algorithms that have not been possible on prior generations of hardware. This is not only an interesting capability in and of itself, but also provides an impetus for the community to fundamentally rethink what a quantum algorithm can look like and to go beyond the limitations in current quantum algorithms.

\vspace{0.3cm}

The remainder of this paper is structured as follows. In \autoref{sec:background} we describe the context in which our work is positioned relative to current quantum computing methods.
Section \ref{sec:enabling_new_hybrid_program} introduces the software development methodology that enables this new type of programming and the associated hardware challenges.
In \autoref{sec:applications} we outline two quantum computing algorithms that are made possible with this new capability.
Finally, in \autoref{sec:hardware_results}, we present an early implementation of one of the algorithms on a physical hardware device and review the results of its execution and associated trade-offs.

\begin{figure*}[ht!]
    \includegraphics[width=0.95\textwidth]{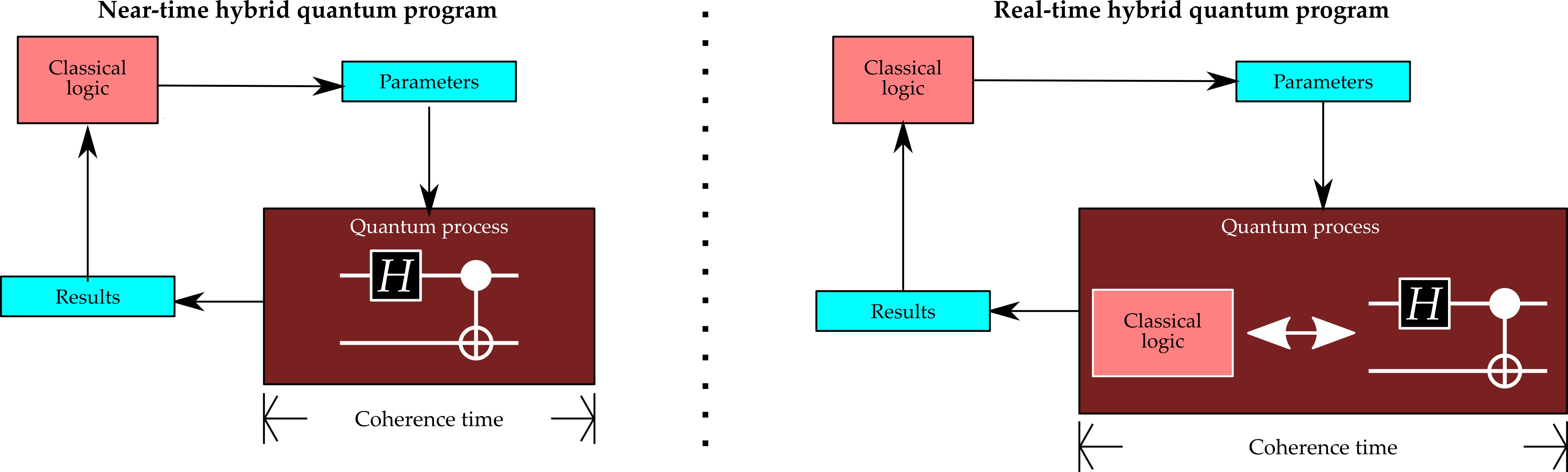}
    \caption{
        A schematic comparison between near-time hybrid programs (left) and the real-time hybrid quantum programs considered and implemented in this work (right). Importantly, in the more advanced form we consider here, hybrid quantum--classical programs can make classical decisions based on the results of quantum measurements, and then use those decisions to condition and control future quantum operations within the coherence times of quantum registers.
    }
    \label{fig:hybrid-comparison}
\end{figure*}


\section{Background}
\label{sec:background}

We review here prior efforts on which our work is based.
First, we examine characteristics of an established hybrid approach used in many algorithms available to users of today's quantum computers.
We then consider algorithmic enhancements that take advantage of recent hardware advances such as mid-circuit measurement and reset along with near-term implementations of real-time classical computation.
Taken together, these features of the current and upcoming generations of hardware describe the current state of existing hybrid quantum/classical computation.


\subsection{Hybrid Quantum Applications}
\label{sec:hybrid_quantum_algorithms}

We first review two essential hybrid algorithms that interleave classical and quantum processing to reduce resources such as overall circuit size, depth, and number of qubits as a way to  work around the limited coherence time and fidelity of qubits today.
However, the constraints of data transfer between classical and quantum processors pose a barrier to leveraging such schemes on sufficiently large devices. Below, we highlight several specific challenges for practical use of these hybrid algorithms.

\vspace{0.2cm}
\paragraph{Variational Quantum Eigensolver (VQE)}

Estimating ground (and excited) state energies with accuracy $\varepsilon$ is at the core of many quantum applications in chemistry \cite{Yuan20_015003, Troyer17_7555, Troyer21_033055} and materials science  \cite{bauerhybridmaterials2016}.
No efficient classical algorithms are known that run in time $\operatorname{poly}(\log(1/\varepsilon))$, whereas a quantum computer makes this possible in principle.
With VQE, a Hamiltonian describing a physical system is simulated using iterative execution of a quantum circuit which prepares an approximate wavefunction, the so-called ``ansatz,'' and a variational algorithm is used to upper bound its ground state energy \cite{peruzzoVariationalEigenvalueSolver2014}. 
By suitably grouping Hamiltonian terms, the family of quantum operations $U(\theta)$ used in the ansatz and the circuit that simulates Hamiltonian terms can be chosen to have a low number of entangling gates, which is advantageous for execution on near-term quantum devices.

VQE is inherently a hybrid algorithm, as classical code is used to optimize the parameter vector $\theta$ in the ansatz $U(\theta)$ by applying a classical method such as gradient descent, SPSA \cite{spallOverviewSimultaneousPerturbation1998}, or the gradient-free Nelder--Mead method, while the quantum computer is used to evaluate a cost function (the sum of estimated Pauli expectations provides an approximate upper bound of the minimum energy).
Classical computations alternate with quantum computations so that the quantum state does not need to stay coherent while classical optimization takes place.
An advantage of this hybrid method over other quantum algorithms is the trade-off of circuit size against total number of repetitions necessary to reach a target accuracy of $\varepsilon$.

\vspace{0.2cm}
\paragraph{Quantum Approximate Optimization Algorithms (QAOA)}

Optimization problems such as \textsc{max-cut} are among the most important tasks addressed by classical methods.
In order to find solutions more efficiently with quantum computing, we can employ the QAOA algorithm \cite{farhi2014quantum} by expressing optimization problems in terms of finding the highest energy configuration of a spin Hamiltonian that is diagonal in the computational basis.
A variational algorithm is then used to optimize, with suitable classical optimization methods, weights applied to a quantum circuit consisting of alternating evolution under the problem Hamiltonian and a non-diagonal mixing Hamiltonian used to move amplitude between different configurations.
An additional parameter defines the number of rounds applied in the QAOA scheme and impacts depth of the resulting circuit.
For a large number of rounds, the Trotter--Suzuki decomposition is a limiting case, while for a small number of rounds the method is not more powerful than classical methods \cite{hastings2019}. 

Similar to VQE, this algorithm is hybrid as the classical code used to optimize the parameter vectors is interleaved with quantum processing used to evaluate the cost function.
The quantum state need not stay coherent while classical optimization takes place.
An advantage of this hybrid method over approaches based on the Trotter--Suzuki decomposition is to trade off depth of circuit size with quality of solution.

\vspace{0.3cm}
A common structural element of both of VQE and QAOA is the alternating execution of classical optimization code with quantum code. While this specific structure makes it possible to perform these algorithms on the current generation of quantum computers, it introduces significant challenges in minimizing the total time to solution.
Several solutions have emerged to partially address these.

A non-negligible time is required to compose a quantum circuit and modify its parameters prior to each iteration.
Some quantum software frameworks \cite{qiskit_advanced_circuit_tutorial,cirq_basics,cirq_rigetti_devices} support "parameter" values as arguments to quantum operations. 
A quantum circuit may be created using parameters instead of fixed values and is compiled once with these symbolic values. Prior to each execution, current parameter values are injected by a second compiler pass or in the backend system, resulting in a reduction in time consumed by program composition.

Compounding this is latency involved in initiating execution of a quantum circuit and communication between the classical and quantum computers \cite{smithPracticalQuantumInstruction2016}, especially pronounced when using a cloud computing service.
Rigetti's quantum--classical cloud platform was one of the first systems to offer execution of classical code on a system physically co-located with the quantum system, considerably reducing data transfer latency \cite{Karalekas_2020}.
The Qiskit Runtime system \cite{johnson_faro_2021,qiskit_runtime} and Quantum Serverless \cite{quantum_serverless} take this a step farther, co-locating the classical and quantum systems, but also executing all iterations of an iterative application as a single job, eliminating lengthy queuing times introduced when many users share a single system.

Both of these approaches offer faster near-time execution, albeit not real-time execution of classical code while the quantum state remains coherent (terminology introduced by IBM Research \cite{quantum_realtime}); see \autoref{fig:hybrid-comparison} for a schematic comparison between near-time and real-time quantum execution.
A significant latency still remains in the time taken to initialize the quantum circuit and the transfer of data between the classical processor and the quantum processor.
To bring execution times down and get closer to real-time, more advanced methods are needed.


\subsection{Mid-Circuit Measurement}
\label{sec:mid_circuit_measurement}

In a recent innovation, both IBM \cite{nation_johnson_2021,qiskit_mid_circuit_tutorial} and Quantinuum \cite{gaebler2021suppression,moore_2020} have shown ways to perform measurement in the middle of executing a quantum circuit, with several benefits.
One, a measurement can be taken and qubit state reset afterwards, effectively enabling the reuse of qubits.
This facilitates the implementation of algorithms that require fewer qubits by using an ancilla that provides state that is carried forward. An array of the individual measurements can be returned and additional analysis performed by a classical driver program.

Two, each measurement can trigger branching to logically different parts of a circuit. This permits more complex algorithms to be encoded within a single execution of a quantum program.  However, the use of branching alone requires that all subsequent computational paths be delineated in the circuit, either in code or using a lookup table, an approach which can grow exponentially with the number of branch points. While this offers an improvement in program control, the benefit is lost with larger problems.
Nonetheless, there are a number of algorithms, a few described below, that are able take advantage of the mid-circuit measurement capability to implement programs that have lower depth and require fewer qubits.

\vspace{0.2cm}
\paragraph{Repeat-Until-Success methods (RUS)}

\citet{paetznick2013} introduced ``repeat-until-success'' (RUS) circuits, a method that is useful for the $\varepsilon$-approximate synthesis of unitaries over basic gate sets such as the Clifford$+T$ set. 
The targeted unitary is implemented with a probability $p$ which constitutes the ``success'' case and fails with probability $1{-}p$ which results in the given state being affected by a Clifford gate.
This leads to a branching on a classical bit indicating success: if the successful branch is taken, the computation continues and the next gate is applied, if unsuccessful, the affected Clifford gate is undone and another attempt is made for the same gate (up to a maximum since coherence time is limited).
It has been shown \cite{BRS:2015a,BRS:2015b} that RUS circuits can lower the expected cost of approximating $R_z$ rotations from $4 \log_2(\varepsilon)$ for deterministic methods \cite{KMMb12,RoSelinger} to $c \log_2(\varepsilon)$, where the constant $c$ is independent of the rotation angle and $c{\approx}1$. See also \citet{KLM+2022} for a recent overview of synthesis methods, including probabilistic methods such as RUS.  

\vspace{0.2cm}
\paragraph{Other Iterative Algorithms}

Other quantum algorithms are iterative in nature and can benefit from the interleaving of quantum/classical computation, processing of mid-circuit measurements, or both.
For example, iterative phase estimation (IPE) functions with a smaller number of qubits than quantum phase estimation, its non-iterative equivalent. IPE measures and resets the state used to read out phases along the way, relying on classical processing to compute a result. 
The semi-classical Fourier transform differs from a normal quantum Fourier transform in that a result distribution is obtained using a single qubit that undergoes a sequence of Hadamard operations, measurements and phase corrections that depend on previous measurements \cite{Griffiths_1996}. This provides a useful resource tradeoff between the number of qubits required and the classical computation of phase corrections.
Two other examples use hybrid computing at the quantum algorithm level: 1. the quantum sieving algorithm introduced by \citet{Kuperberg:2003} uses hybrid computing to cut down the time-complexity of the dihedral hidden subgroup by re-grouping and further processing quantum registers depending on mid-circuit measurement results. 2. the quantum rejection algorithm \cite{ORR:2013} uses hybrid computing to implement non-unitary operations based on attaching auxiliary qubits followed by mid-circuit measurements and branching on the outcomes. 
In addition, there are phase estimation algorithms recently introduced that use a novel
technique known as the quantum singular value transformation (QSVT) \cite{PRXQuantum.2.040203}.

\vspace{0.3cm}
In all these cases, the quantum circuits make use of mid-circuit measurement and reset to get more out of the quantum processor, re-using qubits without exiting the program or using branching to adapt the circuit to changes in the quantum state.
Combined with interleaving of classical and quantum processes, this represents a second level of hybrid quantum programming.
However, this is only a small step towards taking full advantage of the power of the quantum computer.


\subsection{Real-Time Classical Computation}
\label{sec:real_time_classical_computation}

There is another level of sophistication in hybrid algorithms that is the subject of our work.
\citet{corcoles2021dyncircuits} introduce the concept that an adaptive version of iterative phase estimation, exploiting dynamic circuits, could offer a substantial advantage when noise and latency are low. 
They explore the effect of conditional branching on qubit measurement and selection of a new angle from a lookup table given the measurement result.
This is an important first step, as it is an improved form of branching on measurement data.
However, to achieve ``real-time'' and scalable computation, we need a more efficient solution that does not require programming of all branch paths. 

Ideally, the results of one or more measurements could be used as input to arithmetic operations that influence subsequent processing. Rather than branch to specific hard-coded subsequent operations, the rotation angles applied to gates within the program can be the result of a computation in which the angle values change every time. This is effectively an adaptive quantum program, where variables are modified as the execution progresses.  Each iteration may perform a progressively refined computation that converges to an answer, re-using the set of parameterized quantum operations embedded in the program.

An early example of this capability was demonstrated in a randomized benchmarking (RB) application in \citet{reinhold_2019}.  In that experiment, each shot, or repetition, of the RB circuit was comprised of a different sequence of random gates.  This sequence was generated with an embedded classical co-processor, which was programmed to calculate pseudo random numbers using a linear-feedback shift register algorithm.  These random numbers were then used in real-time to choose which Clifford gate to apply to the qubit.
Such an approach to RB has an enormous advantage over pre-computing the entire sequence of random gates prior to execution \cite{granadeAcceleratedRandomizedBenchmarking2015} and highlights one powerful application of real-time classical computation. 
In another example, \citet{ofek_petrenko_2016} demonstrated the use of real-time feedback as a key component in Quantum Error Correction (QEC).

\vspace{0.3cm}

There are many language and hardware-independent approaches to quantum-classical programming, several of which are reviewed in \citet{smithPracticalQuantumInstruction2016,MCCASKEY2018245}.
One effort underway, relevant to our discussion of hybrid quantum programming, is an enhancement of the popular OpenQASM 2.0 specification for quantum circuit definition, called OpenQASM 3.0 \cite{cross2021openqasm}.
A quantum program defined in OpenQASM 3.0 can include classical computations as part of the definition of a quantum circuit.

Complementing this, the QIR Alliance \cite{qir_alliance} is working to develop a Quantum Intermediate Representation (QIR) that can represent such programs that may involve arbitrary interleaving of quantum and classical computation within a single program. While OpenQASM 3.0 is a human-readable representation of a quantum program, the role of QIR is to provide a format that is optimally machine-manipulable, compatible with many existing languages and compiler tools.
While many of these efforts are early stage, the move to embedding classical computation within a quantum program is an important direction for the future of quantum computing.


\section{Enabling a New Form of Hybrid Program}
\label{sec:enabling_new_hybrid_program}

Advancing quantum/classical computational capabilities requires an end-to-end stack where each layer has a clear and distinct purpose, as well as an arsenal of tools that enables integration with such a stack. 
Ideally, an application program is written in a form that makes it easy for users to represent all elements of the solution concisely and in a portable fashion.
It should also be easy for providers of backend systems to convert the intermediate representation of the solution to execute on specific hardware architectures.
We propose that both can and need to be accomplished by introducing a compilation stage that targets a language and hardware agnostic holistic program representation to a backend specific profile.

Below, we examine the improvements necessary in the software stack to support a comprehensive form of quantum/classical computation \cite{smithPracticalQuantumInstruction2016}.
This includes a discussion
about the proposed Quantum Intermediate Representation and a look at challenges to implementation on backend hardware systems


\subsection{The Quantum Software Stack}
\label{sec:quantum_software_stack}

To our knowledge, applications executed on quantum hardware so far have been limited by the inability to execute classical computations while the quantum state remains coherent.
This is evidenced by the prevalence of algorithms such as VQE and QAOA, outlined in \autoref{sec:hybrid_quantum_algorithms}.
These algorithms consist of an outer loop that alternates classical optimization of parameters and execution of a quantum cost function that uses those parameters, requiring the quantum state to remain persistent only for the duration of an iteration. 
Even so, the practicality of leveraging such algorithms is limited by the added latency due to the required data exchange. 

Practicality can be improved by reducing compilation times using symbolic representation of parameters or minimizing latency by co-locating the classical and quantum processors.
In principle, advanced multi-processor systems could achieve very low-latency communication between the classical computer and the quantum control and readout logic.
However, to go beyond this and enable tightly integrated classical processing within the quantum application requires additional support throughout the entire hardware and software stack.

While several dedicated quantum programming languages have been developed~\cite{s.g.t+_enablingscalablequantum_2018, cross2021openqasm, b.b.g+_silqhighlevelquantum_2020}, the predominant approach within the ecosystem largely relies on leveraging popular classical languages such as Python to generate a quantum circuit~\cite{qiskit_org, pyquil, google_cirq, s.d.c+_ketretargetablecompiler_2020}. Such code generation or metaprogramming frameworks rely on the host language to provide the convenience and expressiveness to concisely and comprehensively articulate the program intent, and can present a comprehensive API to a quantum compiler.
The actual quantum circuit is defined by invoking API calls to build a program abstraction in the form of a data structure. This intermediate data structure can be transformed and optimized by the framework before it ultimately generates native hardware instructions to be executed by the targeted quantum processor, a process that requires a significant amount of logic and sophistication.

Another approach is to enhance the semantics of a high-level language, such as C or Python, with syntax to specify that certain loops, variables, and arithmetic computations are to be executed within the quantum program that is produced. 
Recently, full-featured languages such as \qsharp~and extended program representations such as OpenQASM 3.0 have emerged that embed these constructs directly into the language so that a user is able to program using a unified abstraction and the compiler is able to perform the necessary transformations seamlessly.

With any approach, support for classical processing while qubits remain live requires representing the logic for data exchange and processing within an integrated intermediate program representation.
Whether the quantum application is expressed using a domain specific language or a metaprogramming framework, both its program abstraction and its intermediate representation must not be limited to capturing merely quantum operations but need to include classical computation and control flow as well.


\subsection{Quantum Intermediate Representation}
\label{sec:quantum_intermediate_representation}

\begin{figure*}[t!]
    \begin{minipage}{0.3\textwidth}
        \includegraphics[width=0.95\textwidth]{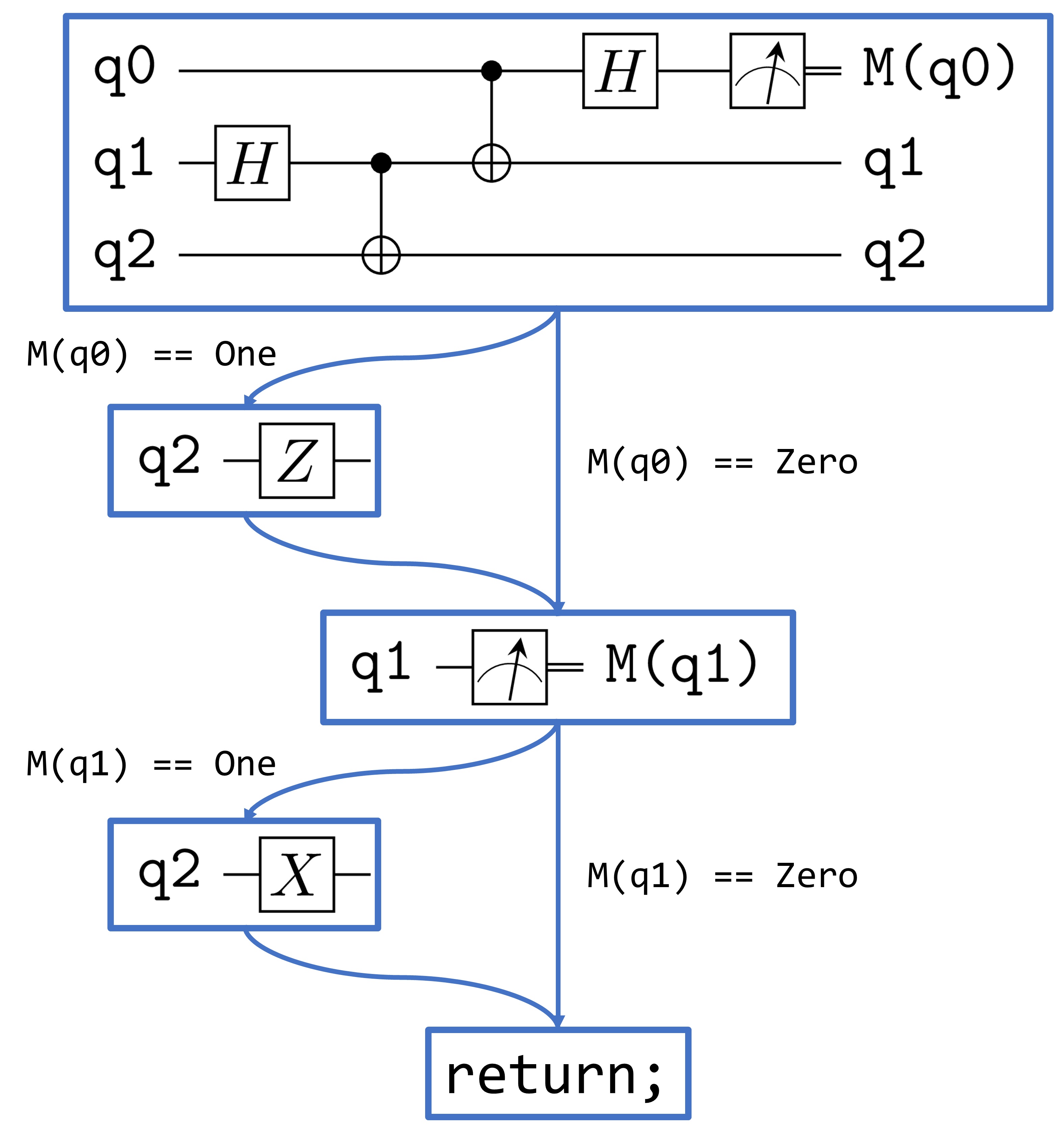}
    \end{minipage}
    \qquad
    \begin{minipage}{0.3\textwidth}
\begin{lstlisting}[
            language=qsharp,
            basicstyle=\ttfamily\tiny
        ]
namespace TeleportationExample {
    open Microsoft.Quantum.Intrinsic;

    @EntryPoint()
    operation RunTeleportation() : Unit {
        use q0 = Qubit();
        use q1 = Qubit();
        use q2 = Qubit();

        // Prepare an entangled pair.
        H(q1);
        CNOT(q1, q2);

        // Perform a Bell measurement.
        CNOT(q0, q1);
        H(q0);

        // Apply X and Z corrections.
        if M(q0) == One {
            Z(q2);
        }
        if M(q1) == One {
            X(q2);
        }
    }

}
\end{lstlisting}
    \end{minipage}
    \qquad
    \begin{minipage}{0.3\textwidth}
        \includegraphics[width=0.95\textwidth]{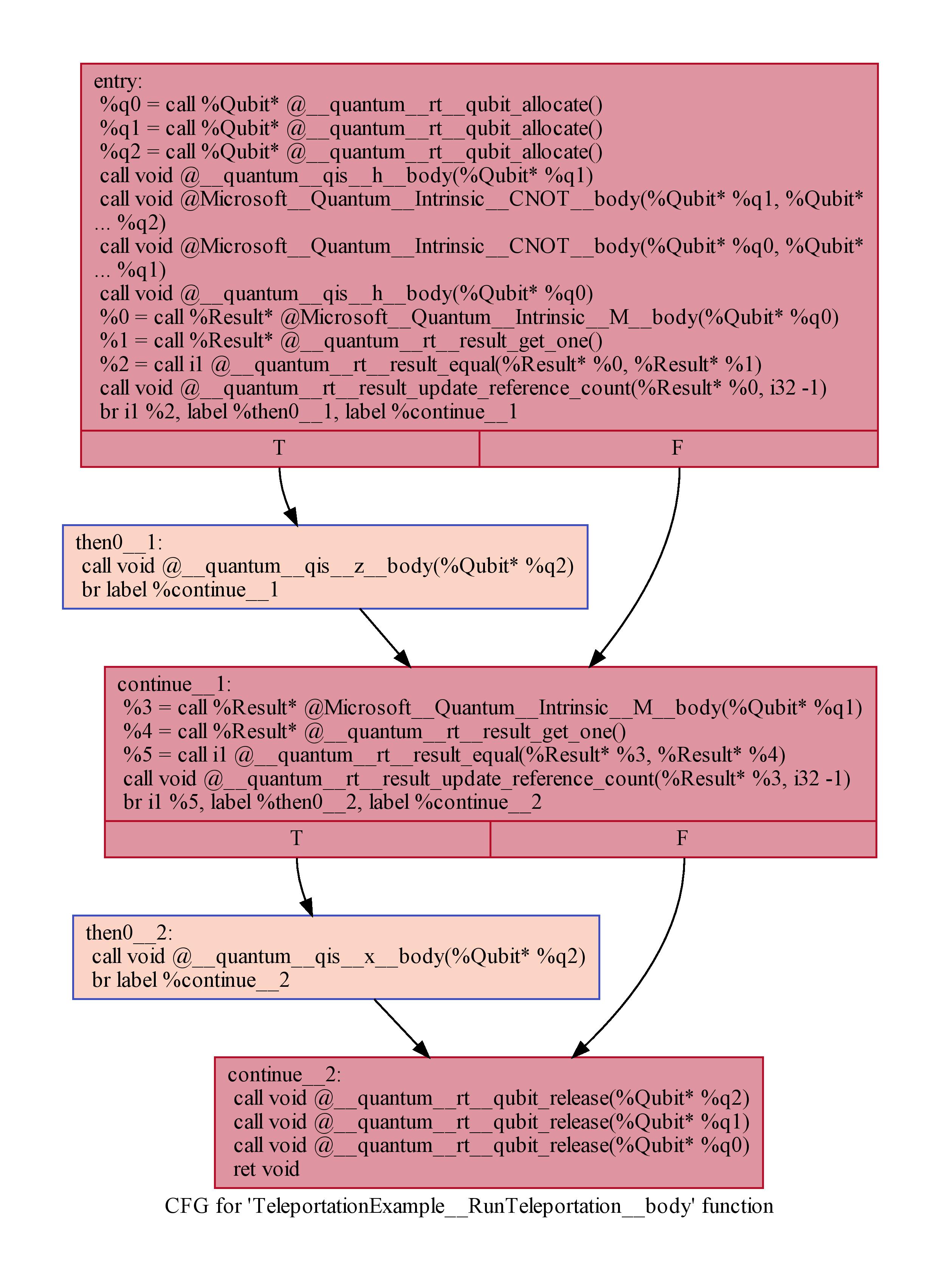}
    \end{minipage}
    \caption{
        An example of compiling a simple \qsharp~program to QIR.
        (left) A schematic representing quantum teleportation as control flow between circuit-like blocks.
        (middle) \qsharp~implementation of quantum teleportation.
        (right) The control flow graph for the QIR generated from the \qsharp~source code. \\
        Note that the conditional branching in the schematic at left and in the control flow graph at right is represented in the control graph as arrows between \emph{blocks} of quantum and classical instructions. Within each block, quantum gates such as \lstinline+__quantum__qis__x__body+ can be called using the \lstinline+call+ instruction in QIR.
    }
    \label{fig:qsharp-to-qir-example}
\end{figure*}

Challenges related to maximizing utility of a dedicated accelerator working in concert with a central processing unit are not unique to quantum computing.
The use of GPUs in modern computing inspired strategies for data exchange between different processors and for facilitating code portability and integration with existing tools and technologies.
Quantum processors, however, are early in their development and to promote and accelerate innovation it is crucial that we do not standardize on a representation for quantum programs that is specific to a particular backend or default to a least common denominator approach to deal with diverse hardware technologies.

To address this challenge in quantum computing, and specifically for the demonstration in this paper, we identified these goals for an effective quantum intermediate representation:

\begin{enumerate}
\item Reduce and accelerate the development effort for software frontends and hardware backends.
\item Permit application and library code to take advantage of novel and unique backend capabilities while maintaining code portability and interoperability. 
\item Enable incremental progress in how different subprocessors or processor components interact and communicate.
\end{enumerate}

A quantum program written in a high-level language is compiled to an intermediate representation that can be executed on a variety of backend systems.
While quantum computing may be unique in many regards, a large part of the required functionality to leverage advanced compilation techniques is not.
To accelerate advancement in quantum, our efforts take advantage of the decades of experience at our disposal on program and dependency analysis, powerful tools for code transformations and optimization, as well as versatile infrastructure for linking and machine code generation.

\vspace{0.3cm}

For these reasons, we chose to build on top of an LLVM-based quantum intermediate representation (QIR).
LLVM is a mature collection of modular and reusable compiler and toolchain technologies~\cite{llvm_project}. QIR is a language and hardware agnostic format that allows for full interoperability between quantum languages and libraries, rewrite steps and optimization passes, and code generation for quantum hardware.
In particular, QIR builds on the design goals of LLVM IR and extends those into the quantum domain:
\begin{quote}
LLVM is a Static Single Assignment (SSA) based representation that provides type safety, low-level operations, flexibility, and \textbf{the capability of representing ‘all’ high-level languages cleanly}. It is the common code representation used throughout all phases of the LLVM compilation strategy. \cite{llvmlanguagereference}
\end{quote}

Using a common intermediate representation allows the software stack to support different source languages and execution platforms without large amounts of redundant development, to keep pace with a significant evolution of the quantum processor architecture over time.
To that end, QIR is an integrated program IR representing not only quantum instructions (e.g.: gate calls and measurements), but classical logic concepts such as branching, memory management and variables.

As an illustration of QIR,
\autoref{fig:qsharp-to-qir-example} shows a brief example of how a simple program such as quantum teleportation can be thought of as a flow through different blocks of quantum instructions with classical branching on measurements (left pane). To the right of this is shown the equivalent \qsharp{}  program and the QIR/LLVM code that is generated. 

QIR specifies how to represent quantum subroutines using a subset of the LLVM IR, following a similar approach as the NVVM compiler IR~\cite{nvvm_ir_spec}, designed to represent GPU compute kernels.
A set of QIR profiles is defined, each of which imposes additional rules that restrict the IR to contain only those constructs that will be executed on a specific QPU target.

Rather than require each frontend language to compile into a processor specific profile, we introduce a compilation stage which maps a QIR program to a targeted QIR profile. An initial implementation for this stage is provided by the Quantum Adaptor Tool (QAT)~\cite{qat_2022} with some custom tooling to map the intermediate representation to a specific backend architecture.
This permits the development of hardware-targeted capabilities without needing to specialize quantum languages to depend on specific features of different devices.

The QIR profile for a specific hardware device defines its quantum gate set, its measurement capabilities, the control flow constructs and classical computations that it can reasonably support. Any program logic that cannot be reduced to leverage only the supported profile will need to be executed as pre- and post-processing steps much like the common practice today. Programs that inherently require a unique hardware feature can execute only on hardware that supports that feature, yet the choice of representation does not add portability constraints to those fundamental to the algorithm. Conversely, a hardware feature that is not represented in the QIR will not be accessible to users. A vital step towards making quantum computing practical will require agreement in the community about the operations supported in a QIR as well as the transformations and optimizations applied at compile time for both the quantum operations and the classical computations within any program.


\subsection{Hardware Challenges}
\label{sec:hardware_challenges}

Any quantum program, written in a user-level programming language, is typically converted to a hardware-specific set of instructions that executes on a quantum computer system.
To enable quantum programs that use a new form of hybrid quantum-classical computation, we must account for the limitations that are inherent in this generation of quantum computing system and consider how these systems may evolve.

The quantum elements (or `qubits') assembled into a quantum computing system are manipulated using classical control electronics to generate sophisticated sequences of microwave or laser pulses depending on the technology used in the system.
The nature of these quantum elements dictates that pulses on them are defined on a nanosecond timescale, and the relative timing of pulses on different quantum elements must be precisely coordinated. Furthermore, to enable any useful control of a qubit, programmable control flow that operates on the same timescale is a requirement. Since a general-purpose CPU is not sufficient for this, a quantum control system is often constructed with specialized hardware such as an arbitrary waveform generator (AWG) or a field programmable gate array (FPGA) that can be utilized to meet these stringent requirements. At scale, a quantum computer is built with many of these. Newer generations are becoming more sophisticated, for example, all on a single chip.

Unsurprisingly, the advantages of special purpose hardware come with loss of general purpose computing features including smaller instruction sets and less runtime memory. While specific tradeoffs may be redressable, we describe here some possible limitations when working with such specialized quantum control systems.

Arithmetic operations performed on such systems may be different from those of a general-purpose processor. To minimize memory usage and execution time, fixed point numerical representations are commonly used: an integer of some number of bits with an implicit decimal point. Other than division, basic arithmetic is fast, the periodic 2’s complement format provides an effective representation of angles, and interpolation may be used to implement operations such as sine and cosine. 
However, fixed point representations are sensitive to both overflow and underflow. When not used to represent periodicity, the numeric range is quite limited, and for example, multiplication by 0.5 results in the loss of one bit of precision. The developer of a hybrid quantum program that includes such arithmetic operations will need to be aware of any constraints imposed by the target hardware system.

There is also a challenge in the generation of hardware level instructions for a hybrid program that includes classical computation. As long as the arguments to quantum gates are known at compile-time, a transpiler can generate hardware-specific code and pulse sequences to perform those operations efficiently. However, to enable variable arguments to quantum gates, the compiler must generate a more complex sequence of code. A common way to implement support for variable arguments on current architectures is to use a \lstinline+RZ+ basis gate since it may be implemented virtually \cite{mckay2017efficient}, i.e., as only a phase change on subsequent pulses instead of as its own pulse. This means it has effectively perfect fidelity and has zero run-time cost. For this gate to be ``virtual'' but still accept variable arguments at run-time, the classical processor must support arithmetic and trigonometry.

There are many other challenges associated with this new form of quantum program that adapts its execution to changes in variable state. 
A program that executes a fixed series of operations returns a dataset with a predictable structure, but if the paths are modified during execution the structure of the return data can vary across executions. 
Another complexity stems from the fact that qubits are often manipulated in parallel, using what is essentially a network of small classical processors, and the bandwidth, latency, and connectivity of such networking could be rate limiting factors at scale. 
In these early stages, the introduction of classical computation to quantum programs will be constrained by these multiple challenges.
Future generations of quantum control systems should take these challenges into account.


\section{The Path to Scalable and Reliable Applications}
\label{sec:applications}

With a quantum software stack that supports classical computation and a portable intermediate representation, we have the ingredients to enable a compelling advance in quantum programming.
Existing hybrid algorithms integrate quantum and classical computation, but in a restricted and disjoint fashion.
We break new ground here by describing a form of adaptive hybrid programming in which quantum and classical computational primitives may be tightly interwoven, rendering optional the need for quantum measurement data to be repeatedly transferred across computer interfaces.

This capability inspires development of an entirely new class of quantum algorithm, one in which reliability can be improved and the breadth of applications extended.
We illustrate this idea with an example of a simple quantum algorithm that uses both quantum and classical operations, followed by a discussion of the range of features essential to making this new capability complete.  Later, we describe the workings of an advanced algorithm that we execute on both a quantum simulator and on quantum hardware (results presented in \autoref{sec:hardware_results}).


\subsection{Resetting a Quantum System}
\label{sec:reset_quantum_system}

To show how classical computation can be used to enhance a quantum program, we highlight an algorithm designed to reset a quantum bit from an unknown state in the shortest amount of time required to achieve a desired ``fidelity''. The probabilistic nature of a quantum state makes it challenging to implement a quantum reset protocol that has 100\% certainty of success in a single operation \cite{navascues_2018}. Several tutorial examples \cite{ibm_quantum_cond_reset,ella_2022} demonstrate how multiple qubit reset operations are required to achieve a high probability of success and how the number used can affect the fidelity of the operation.

While this example is primitive, it serves to highlight the use of program variables, classical loop execution, and simple arithmetic during the time domain of quantum program execution. Algorithm \ref{alg:qubit_reset} describes its program logic.

\begin{algorithm}[h!]
    \caption{Active Qubit Reset}
    \label{alg:qubit_reset}
    \begin{algorithmic}[1]
    \State $req\_successes \gets 2$
    \State $num\_successes \gets 0$
    \For{$counter \gets 0, 4$}
        \State $value \gets measure(qubit)$     \Comment{measure $qubit$}
        \If{$value == 0$}
            \State $num\_successes \gets num\_successes + 1$
        \Else
            \State $X(qubit)$               \Comment{flip $qubit$ state}
            \State $num\_successes \gets 0$
        \EndIf
        \If{$num\_successes == req\_successes$}
            \State $break$      \Comment{done, exit loop}
        \EndIf
    \EndFor
  \end{algorithmic}
\end{algorithm}

The algorithm succeeds when it measures 0 twice in a row on the qubit. If not successful after 4 measurements, the loop exits and the reset has failed (not shown). Multiple variables are defined and used within a classically controlled loop and a counter is updated using simple classical arithmetic.
Operations such as these can be constructed using various hardware-specific libraries \cite{ella_2022} that support classical operations and pulse control in the same program.
However, OpenQASM-level APIs used to develop portable algorithms and applications typically permit branching on mid-circuit measurement but little else that is classical in nature (Section \ref {sec:mid_circuit_measurement}).
Our work is specifically targeting general solutions with open specifications, available for implementation across multiple target platforms.

\vspace{0.3cm}

With classical computation embedded in the quantum program, flow control logic and values applied to gate operations may be computed ``on-the-fly.'' Performing an arithmetic computation based on captured measurement values to modify the current variable values results in a program is ``adaptive,'' i.e. it changes its conditional behavior in response to measurements of a continually changing quantum state.


\subsection{Scaling Up the Software Stack}
\label{sec:scaling_stack}

Several enhancements to the software stack and target hardware are required to enable this new form of hybrid programming. To illustrate, we focus on a particular model of integrated classical and quantum computation: a parameterized series of quantum operations executed repeatedly, interleaved with classical computation of the next set of variable values and control flow, with some iterations occurring while the qubits are kept coherent and the quantum state maintained. This is sufficient to allow us to run algorithms such as random walk phase estimation, described in Section \ref{sec:random_walk_phase_estimation} below.

To accomplish this, our implementation of a quantum/classical hybrid program supports the following programming constructs in addition to features already available:

\vspace{0.2cm}

\paragraph{Parameters and Variables}

In Section \ref{sec:hybrid_quantum_algorithms}, we showed how existing hybrid algorithms minimize latency of circuit composition through the use of parameterization, by which angles used in quantum gates are defined symbolically and the actual values supplied at the time of execution. In these algorithms, the classical values provided at initialization, the ``parameters,'' are used as constants and not modified during execution.

In our advanced real-time hybrid programs, we support classical values that may be changed during the course of execution. These values, or ``variables,'' may be used as rotation angles in some systems, but could also be used as a loop counter or as computed values to be returned to a calling program. The quantum firmware and hardware is sophisticated enough to adjust the execution of the circuit to a new variable value, and to perform this adjustment during the course of its execution. Optimal performance can be achieved if the quantum program does not need to exit in order for new values to be provided.

\vspace{0.2cm}
\paragraph{Variable Arithmetic}

Computing new variable values within the quantum program requires sufficient arithmetic computational capability in the quantum firmware to calculate new values as a function of mid-circuit measurement results or other variables, between execution of quantum operations.
The calculation, including the routing of the measurement results, must happen quickly enough so that the qubits don't decohere in the meantime.

In current hybrid algorithms, the classical code used to execute a quantum circuit and iteratively converge on a solution is typically implemented in a high-level language like Python, C\#, or Julia.
This will not work for arithmetic operations that are to be executed within the context and time domain of the quantum program.
Instead, these instructions need to be converted to low-level assembly or bit codes that can execute in the control system (FPGA or other) and must be synchronized with quantum code execution, requiring advanced control features not exposed in most of today's quantum computers.

\vspace{0.2cm}
\paragraph{Conditional Looping}

Running the parameterized circuit repeatedly -- looping -- requires support in the high-level quantum software for defining the body of the loop (which may include quantum and classical parts) and specifying the loop exit condition (either a variable or a direct measurement).
The upper levels of the software stack, such as the compiler and the intermediate representation, must be capable of recognizing variables, loop constructs, and arithmetic computations that will be executed within the quantum control system, and generating the associated low-level firmware instructions in the assembly language specific to a target hardware system.

\vspace{0.2cm}

Taken together, these features make possible a new class of hybrid quantum/classical program that can exploit the full potential of both types of computing hardware working in tandem.
The integrated quantum program used in our demonstration below was implemented in the Q\# programming language \cite{microsoft_qsharp} and made use of the Quantum Intermediate Representation (QIR) described in Section \ref{sec:quantum_intermediate_representation}. The Q\# compiler follows the approach described in QAT documentation \cite{qat_2022} to separate the classical and quantum portions of the QIR and generate the instructions required by the backend hardware. These represent the first set of programming tools capable of representing this advanced form of hybrid computation.


\subsection{Random Walk Phase Estimation}
\label{sec:random_walk_phase_estimation}

\begin{figure}[t!]
    \begin{tikzpicture}[
    every node/.append style = {draw, anchor = west},
    grow via three points={one child at (0.175,-0.6) and two children at (0.175,-0.6) and (0.175,-1.2)},
    edge from parent path={(\tikzparentnode\tikzparentanchor) |- (\tikzchildnode\tikzchildanchor)}]
        \node{Phase estimation}
            child {node {Quantum phase estimation \cite{kitaevay_quantummeasurementsabelian_1995}}}
            child {node {Iterative phase estimation}
                child {node {Kitaev phase estimation}}
                child {node {Faster phase estimation \cite{svoreFasterPhaseEstimation2013}}}
                child {node {Robust phase estimation \cite{kimmelRobustCalibrationUniversal2015}}}
                child {node {Adaptive phase estimation}
                    child {node {Dynamic PE \cite{corcoles2021dyncircuits}}}
                    child {node {Particle filter \cite{granadeRobustOnlineHamiltonian2012}}}
                    child {node {RejF PE \cite{wiebeEfficientBayesianPhase2016}}}
                    child {node [red, thick, dashed] {RWPE \cite{rwpe_theory}}}
                }
            };
    \end{tikzpicture}
    \caption{A taxonomy of various phase estimation algorithms.
    }
    \label{fig:pe-taxonomy}
\end{figure}
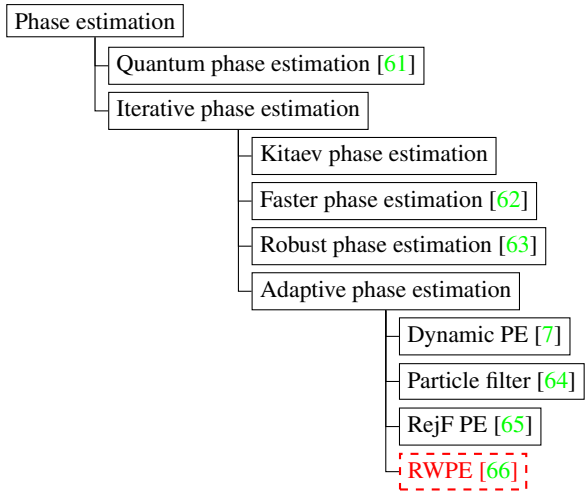

For our demonstration, we focus on the example problem of quantum phase estimation. A variety of algorithms are available to determine the phase inherent in a given quantum operation, as shown in the summary in \autoref{fig:pe-taxonomy}.
We contrast several of these algorithms with our approach, the RWPE algorithm, that uses the new form of hybrid quantum/classical computation introduced in this paper.

Consider the problem in which have a particular quantum subroutine \lstinline|U| whose action is represented by the unitary $U$ whose eigenvalues we would like to learn. For instance, in quantum chemistry, \lstinline|U| may be a step in an algorithm to simulate the Hamiltonian of a given chemical system such that the eigenvalues of $U$ represent energy levels of that system.

In the case that $U = e^{i H \tau}$ for some Hamiltonian of interest $H$ and for some time interval $\tau$,  then one can na\"ively approach the problem of finding the minimum eigenvalue $E_0$ of $H$ by rephrasing as a minimization problem
\begin{equation}
    \label{eq:vqe-problem}
    E_0 = \min_{\vec{x}} \left\langle \psi(\vec{x}) | H | \psi(\vec{x}) \right\rangle,
\end{equation}
for some parameterized set of state preparations $\ket{\vec{x}}$ known as an \emph{ansatz}. Approximating this expectation from measurements of a quantum device yields the variational quantum eigensolver (VQE) algorithm \cite{peruzzoVariationalEigenvalueSolver2014} described in \autoref{sec:hybrid_quantum_algorithms}.

VQE has a number of limitations that make it difficult to apply for larger problems. In particular, the state preparation ansatz must be called repeatedly as the measurement implied by \eqref{eq:vqe-problem} consumes $O(1 / \epsilon^2)$ copies of the state prepared by the ansatz operation at each iteration in order to reach an accuracy of $\epsilon$.
\footnote{This is completely general for \emph{any} algorithm that uses independent measurements of a quantum state, and follows immediately from that the Fisher information for independent measurements simply adds. Thus, by the Cram\`er--Rao bound \cite{coverElementsInformationTheory2006}, if a single VQE measurement yields Fisher information $I_0$, the variance after $N$ shots is bounded by $\sigma^2 \ge 1 / {N I_0}$, such that $1 / \epsilon^2$ shots are required to ensure that $\sigma \le \epsilon$.}
On the other hand, if we can prepare a register of qubits in an eigenstate $\ket{\phi}$ of $U$ with eigenvalue $e^{i\phi}$, such as by using adiabatic state preparation, then we can use phase estimation to learn $\phi$.

In particular, quantum phase estimation (QPE) uses the inverse quantum Fourier transform (QFT) to prepare a register in the state $\ket{b_0 b_1 \dots b_{n-1}}$ where $\phi = 0.b_0 b_1 \dots b_{n-1}$ is an expression of $\phi$ as a fixed-point binary number using $n$ classical bits \cite{kitaevay_quantummeasurementsabelian_1995}.
To avoid the need for an additional register of $n$ auxiliary qubits, iterative phase estimation (IPE) methods improve upon QPE by estimating $\phi$ using one classical bit at a time \cite{kitaevQuantumComputationsAlgorithms1997,svoreFasterPhaseEstimation2013,kimmelRobustCalibrationUniversal2015}.
Critically, the ideal action of each IPE measurement leaves the eigenstate $\ket{\phi}$ invariant, such that initial state preparation can be reused between iterations up to the extent allowed by noise.

\begin{algorithm}[H]
    \caption{Iterative PE (single iteration)}
    \label{alg:iterative-pe-single}
    
    \begin{algorithmic}[1]
    \Require \text{\texttt{target} must be in an eigenstate} $\ket{\phi}$ \text{of \texttt{U}}
    \Procedure {IterativePEStep}{\texttt{U}, \texttt{t}, $\phi_{\text{inv}}$, \texttt{target}}
        \State \texttt{q} $\gets$ \text{fresh qubit}
        \State \texttt{H}(\texttt{q})
        \State \texttt{Rz}($-\phi_{\text{inv}} \cdot {}$ \texttt{t}, \texttt{q})
        \State \texttt{Controlled(q) U}(\texttt{t}, \texttt{target})
        \linecomment{\texttt{q} should now be in the state $(\ket{0} + e^{i(\phi - \phi_{\text{inv}}) t} \ket{1}) / \sqrt{2}$.}
        \State \texttt{H}(\texttt{q})
        \State \Return \text{the result of measuring \texttt{q} in the} $Z$\text{-basis}
    \EndProcedure
   \end{algorithmic}
\end{algorithm}

By contrast with VQE, both QPE and IPE use $O(1 / \epsilon)$ time to estimate $\phi$, roughly corresponding to the difference between the standard quantum and Heisenberg limits for metrology. This quadratic advantage together with the ability to reuse state preparations allows for PE methods to be much more efficient at estimating eigenvalues than VQE.

In practice, however, noise in near- and medium-term devices can make running IPE challenging due to the large gate depths introduced by calling \lstinline|U| for a variety of different evolution times while qubits remain coherent. To mitigate this, one can consider resetting the eigenstate register when needed; making this decision online, however, requires an online estimate $\hat{\phi}$ of $\phi$ to detect when inconsistent measurement results are observed. The problem of estimating a parameter conditioned on a partial data record is a natural fit for Bayesian inference \cite{granadeRobustOnlineHamiltonian2012}, such that we consider Bayesian PE methods in this section. Adopting a Bayesian approach also allows extending adaptivity to include online experiment design as well as the heuristics used by dynamic phase estimation \cite{corcoles2021dyncircuits}.

If we consider a single IPE iteration of the form listed in \autoref{alg:iterative-pe-single}, then the probability of getting a 1 at the end of the iteration is given by $\Pr(1 | \phi; t) = \cos^2(\phi t / 2)$. This forms a \emph{likelihood function}, such that we can use Bayes' rule to compute $\Pr(\phi | d_0, d_1, \dots, d_{n - 1})$ from a sequence of measurements $\{d_0, d_1, \dots, d_{n - 1}\}$ collected at evolution times $\{t_0, t_1, \dots, t_{n - 1}\}$ \footnote{This expression of the likelihood function also allows for the calculation of Cram\`er--Rao bounds for IPE, as in the work of \citet{ferrieHowBestSample2013}, confirming that IPE requires $O(1 / \epsilon)$ time to reach an accuracy of $\epsilon$.}. The expectation value over this distribution then minimizes the error in our online estimate of $\phi$ \cite{banerjeeOptimalityConditionalExpectation2005}.

Performing exact Bayesian inference can be prohibitively expensive, however, especially within qubit lifetimes. Online approximation methods such as particle filtering \cite{granadeRobustOnlineHamiltonian2012,doucetTutorialParticleFiltering2011} can reduce Bayesian inference to a Markov chain conditioned on experimental measurements, diminishing the cost for Bayesian PE. Rejection filtering phase estimation (RejF PE)  \cite{wiebeEfficientBayesianPhase2016} further reduces costs by using rejection sampling to implement a reduced form of particle filtering, allowing adaptive resets to be performed with fewer classical arithmetic operations.

Recently, the random walk phase estimation (RWPE) algorithm \cite{rwpe_theory}~was introduced to allow for computing online estimates using only a few arithmetic expressions per iteration, making it practical to use PE methods that are iterative on near- and medium-term devices subject to noise.

\begin{algorithm}[H]
    \caption{\label{alg:rwpe-no-unwinding}
        Basic random walk phase estimation algorithm of \citet{rwpe_theory}.
    }
    \begin{algorithmic}        
        \Function{RandomWalkPhaseEst}{$\mu_0$, $\sigma_0$, \texttt{U}}
            \State $\mu_0$: initial mean
            \State $\sigma_0$: initial standard deviation
            \State \texttt{U}: oracle whose eigenvalues are to be estimated
            \State \texttt{target}: A register of qubits prepared in an eigenstate of \texttt{U}.
            \seccomment{Initialization}
            \State $\mu \gets \mu_0$
            \State $\sigma \gets \sigma_0$
            \seccomment{Main body}
            \For{$i_{\text{exp}} \in \{0, 1, \dots n_{\text{exp}} - 1\}$}
                \State $\phi_{\text{inv}} \gets \mu - \pi \sigma / 2$
                \State $t \gets 1 / \sigma$
                \State Sample $d$ from \Call{IterativePEStep}{\texttt{U}, $t$, $\phi_{\text{inv}}$, \texttt{target}}
                \linecomment{$\Pr(d = 0 | \phi; \phi_{\text{inv}}, t) = \cos^2(t (\phi - \phi_{\text{inv}}) / 2)$.}
                \If{$d = 0$}
                    \State $\mu \gets \mu + \sigma / \sqrt{e}$
                \Else
                    \State $\mu \gets \mu - \sigma / \sqrt{e}$
                \EndIf
                \State $\sigma \gets \sigma \sqrt{(e - 1) / e}$
            \EndFor
            \seccomment{Final estimate}
            \State \Return $\hat{\phi} \gets \mu$
        \EndFunction
    \end{algorithmic}
\end{algorithm}

Critical to the execution of RWPE is that the update of $\mu$ and $\sigma$ happens during execution so that the new values of each variable can be used as inputs to \lstinline+U+. This requires us to not only branch based on measurement outcomes, but to maintain a continually changing program state without returning to a remote classical processor.
This is the advantage of the hybrid, adaptive algorithm of this type.

In contrast, to implement the equivalent program logic with existing methods (no classical computation inside the quantum code), one would re-write \autoref{alg:rwpe-no-unwinding} to use lookup tables rather than floating-point variables --- in particular, if we know the whole history of quantum measurements made throughout an RWPE run, then we can reconstruct $\mu$ and $\sigma$. This table grows exponentially in size with the number of measurements made, however. Practical applications may require between 20 and 60 iterations of RWPE (yielding respective relative accuracies of $10^{-2}$ and $10^{-6}$), requiring the storage of prohibitively large lookup tables.

The RWPE algorithm is a clear example of the methods available to a developer with access to this new form of hybrid and adaptive quantum programming.
In the following section, our efforts were focused on demonstrating the practical viability of the approach by executing the RWPE program on a specific quantum computing system.


\section{Execution on Quantum Computing Systems}
\label{sec:hardware_results}

For this work, we executed the Random Walk Phase Estimation program in \autoref{sec:random_walk_phase_estimation} on both a quantum simulator and a physical quantum computing system.
We discuss how this was accomplished, focusing on the interpretation of the intermediate representation RWPE program and the specific parameters used in the program to control its execution.
Execution of this program was performed on a quantum simulator enhanced with classical computation capability and a next-generation quantum computing system provided by Quantum Circuits Inc. (QCI) \cite{qci_azure_2021,qci_website}. The QCI system is one of the first quantum computers designed with a control system that provides the novel capabilities integral to the enhanced hybrid quantum-classical programming model and necessary for execution of the RWPE program that was selected as the primary example.


\subsection{Compiling RWPE and Interpreting QIR}
\label{sec:hardware_1}

The quantum intermediate representation (QIR) discussed in \autoref{sec:quantum_intermediate_representation} enables the abstract definition of an advanced quantum program in a form that is independent of any specific target system.
The QIR can be produced from a variety of higher-level programming languages.  
For this demonstration, we chose to define the RWPE program in the Q\# language (source code shown in \autoref{apdx:source-code}), taking advantage of QIR generation support provided by the \qsharp~compiler.
We then utilized the QAT tool \cite{qat_2022} to apply transformations during compilation to produce QIR compatible with hardware (e.g. by assigning static qubit indices).
Submission and execution of the QIR program is all managed within the Azure Quantum service \cite{microsoft_qsharp}.

The RWPE program in its targeted QIR form can be transformed to the native program representation required for execution on a specific backend quantum computing system.
QCI used a pre-release version of the PyQIR package for Python \cite{pyqir_2022} to parse the QIR representation of RWPE and to perform the transformation within the QCI quantum program compiler.
The mapping from QIR program features to QCI's intermediate representation was mostly a one-to-one translation, with a few exceptions. Specifically, the quantum gates that are generated after transpilation are unique to the QCI hardware and the behavior of division with respect to numeric data types is limited as described below. 

Integer and boolean values are mapped to 18 bit signed integers, while doubles map to Q2.16 fixed-point integers (with range $\left[-2, 2-2^{-16}\right]$). For a numerical value used as an angle, we assume a convention that its value is in units of $\pi$ for a range of 2 full periods.
Native addition, subtraction, and multiplication are supported, but division is implemented with an approximation using interpolation table techniques.
Control flow is implemented with hardware level \lstinline+if+ and \lstinline+goto+ statements
and quantum gates are transpiled to the native gate set, described in \autoref{sec:execution_on_hardware}, using internal methods.
The system has the ability to return intermediate values, including measurements, to the user.

The QCI compiler will raise an error if a variable is set to a value outside its allowed range by the QIR, although there is no run-time check on hardware if a value underflows or overflows. Indeed, in this RWPE program, \lstinline+sigma+ will underflow after about 20 iterations.
The program variable \lstinline+1/sigma+ is used as an angle and its run-time overflow behavior of wrapping around the bounds is acceptable. 
These behaviors were validated for RWPE using simulations.
For many problems of interest, sufficient accuracy may be obtained with these constraints.
 
These limitations highlight two consequences of using a quantum intermediate representation.
First, data types provided in hardware may not match exactly what is specified in the QIR representation of the program. Second, since the same QIR may be used with different hardware, the end user does not need to change their algorithm in order to target different backend systems.
For this reason, it is important for the user to have available a simulator subject to the same classical limitations as the hardware in order to validate program behavior prior to scaling the program up to larger numbers of qubits. See \autoref{sec:execution_on_simulator} for details about QCI's simulator.


\subsection{Parameterizing RWPE Program for Execution}
\label{sec:parameterizing_for_execution}

When a program such as RWPE is executed, a user may want to select options to analyze variations in the run-time behavior.
In this case, the RWPE program shown in \autoref{apdx:source-code} defines an oracle, $U(t) = R_z(-0.5 t)$, for which $\phi = \pm 0.5$ (in units of $\pi$) are the true eigenvalues. The RWPE algorithm will preferentially converge on the eigenvalue at $0.5$ because it is closer to the initial value for the prior estimate \lstinline+mu+.
For this demonstration, the parameter value $-0.5$ to the $R_z$ gate is hard-coded in the program source. Alternatively, this could have been passed as a program parameter, but this was not in place at the time this program was executed. Consequently, the program is always expected to produce the result $\hat{\phi} \approx 0.5$.

For each program execution, the inner code loop is executed $nIter = 24$ times, as the algorithm can make no progress beyond this due to underflow.
Throughout this paper, we refer to such executions as \emph{shots}, and use the term to include all embedded loops and intermediate measurements.
In this case, each shot returns a single datum, the estimated value of \lstinline+mu+. 
Since the eigenstate will decohere across iterations within a single shot, the program includes a parameter specifying a subset of iterations to run before the eigenstate is refreshed (reset and prepared anew).
In general, if the eigenstate is expensive to prepare, it should be refreshed less often as each refresh will increase execution time; the faster the eigenstate decoheres, the more often it should be refreshed to improve accuracy.
In this program, the eigenstate is simply $\ket{1}$, prepared with a reset and an \lstinline+X+ gate, which favors a frequent refresh. For our demonstration, we chose to refresh every other iteration to highlight the existence of this user-selectable trade-off.

The algorithm required a final calculation of \lstinline+mu * 2+, which we did in post processing instead of in the \qsharp~program, meaning the range of resultant values of \lstinline+mu+ was effectively doubled relative to the range of the fixed-point data type. We could obtain a better estimate of the eigenvalue by refitting all measurements as a post-processing step as described in \autoref{sec:future_work}, but we do not do that here. Instead we only show the result estimated at run-time by the program.


\subsection{Execution on an Enhanced Quantum Simulator}
\label{sec:execution_on_simulator}

\begin{figure}[t!]
    \includegraphics[width=0.5\textwidth]{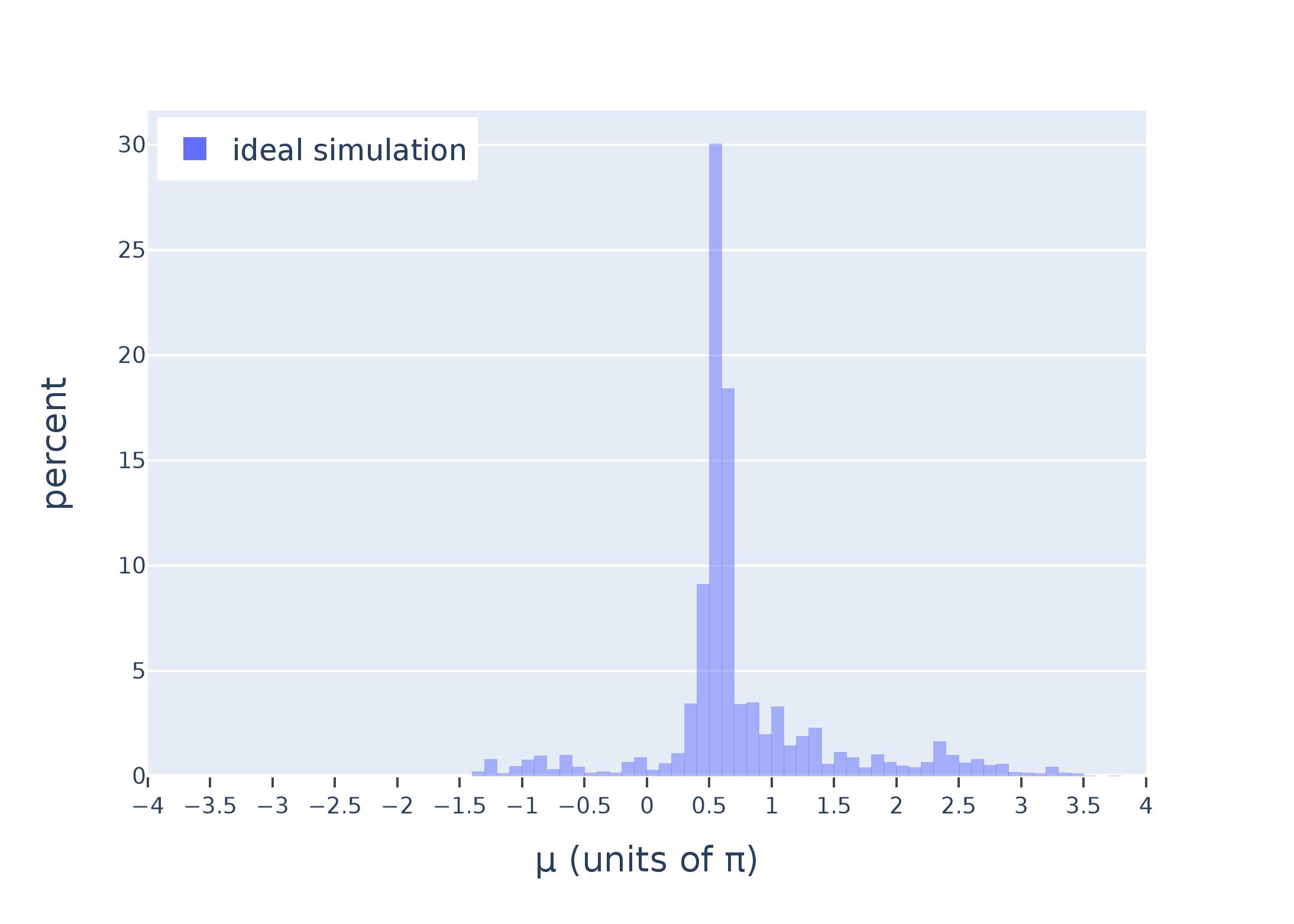}
    \caption{Ideal simulation of 10,000 shots, each resulting in one value of $\mu$ included in this histogram. Each value of $\mu$ shows where the RWPE algorithm ended after 24 iterations, and the success of the RWPE algorithm is exhibited in the highest peak appearing at the eigenvalue $\mu=0.5$. The other values in the histogram highlight the random walk aspect of the RWPE algorithm and show that the algorithm can "fail" even if perfectly executed. The asymmetry of the distribution reflects the starting value, $\mu_0$. The height of the peak depends on the number of bins, which was 100 in this paper.}
    \label{fig:ideal_sim}
\end{figure}

Prior to executing on quantum hardware, it is important to validate the program by running it on a simulator that mimics closely the computational behavior of the target system. 
Existing simulators that execute only quantum operations or which don't match the capabilities of the target hardware are not adequate.

To address this, QCI developed a custom simulator in which control-flow and classical operations are executed in Python and quantum operations on a statevector or a density matrix from Qiskit Terra's
\cite{qiskit_org} \lstinline+quantum_info+ module. Classical registers are simulated using native Python \lstinline+int+ and \lstinline+float+ data types or, to model the hardware closely, the \lstinline+fixedpoint+ Python package. The simulator can also leverage Qiskit Aer's \lstinline+noise_model+ module to include quantum gate and readout noise. Three sources of infidelity can be modeled: a finite number of shots, quantum noise, and hardware-specific classical computation.

We executed the RWPE program on the simulator with no noise configured, i.e. an ``ideal simulation,'' to validate that the algorithm performs as designed.
Results are shown in \autoref{fig:ideal_sim}. To mimic an ideal quantum computer, the statevector simulator is seeded with a pseudo random number generator (RNG). It includes no quantum noise, and classical operations use full precision registers. Each repetition, or shot, of the RWPE protocol generates a single value for \lstinline+mu+ corresponding to a different RNG seed. The only source of error in this simulation is associated with the RNG in the finite number of shots executed.
The result of the simulation is a distinct peak in the histogram at the expected eigenvalue of $0.5$.

We then ran the RWPE simulation with a noise model that approximates the characteristics of the target hardware to predict the behavior to be expected when the program is executed on that system. 
Data were obtained from execution of the RWPE program on this "noisy" simulator and compared against results obtained when running on hardware, as discussed in \autoref{sec:execution_on_hardware} below.


\subsection{Execution on Quantum Hardware}
\label{sec:execution_on_hardware}

In this section, we present results obtained from executing the RWPE program on a quantum computing system provided by Quantum Circuits Inc. (QCI) \cite{qci_azure_2021,qci_website}.
For its work with Microsoft Azure, QCI had deployed for testing and validation a quantum computer designed around superconducting 3-D resonator technology and which provides a hardware-efficient platform for the development of advanced quantum algorithms.
The control system that manages the components of this system has been implemented with many of the quantum/classical computational features described in this paper.

\begin{figure}[t!]
    \includegraphics[width=0.5\textwidth]{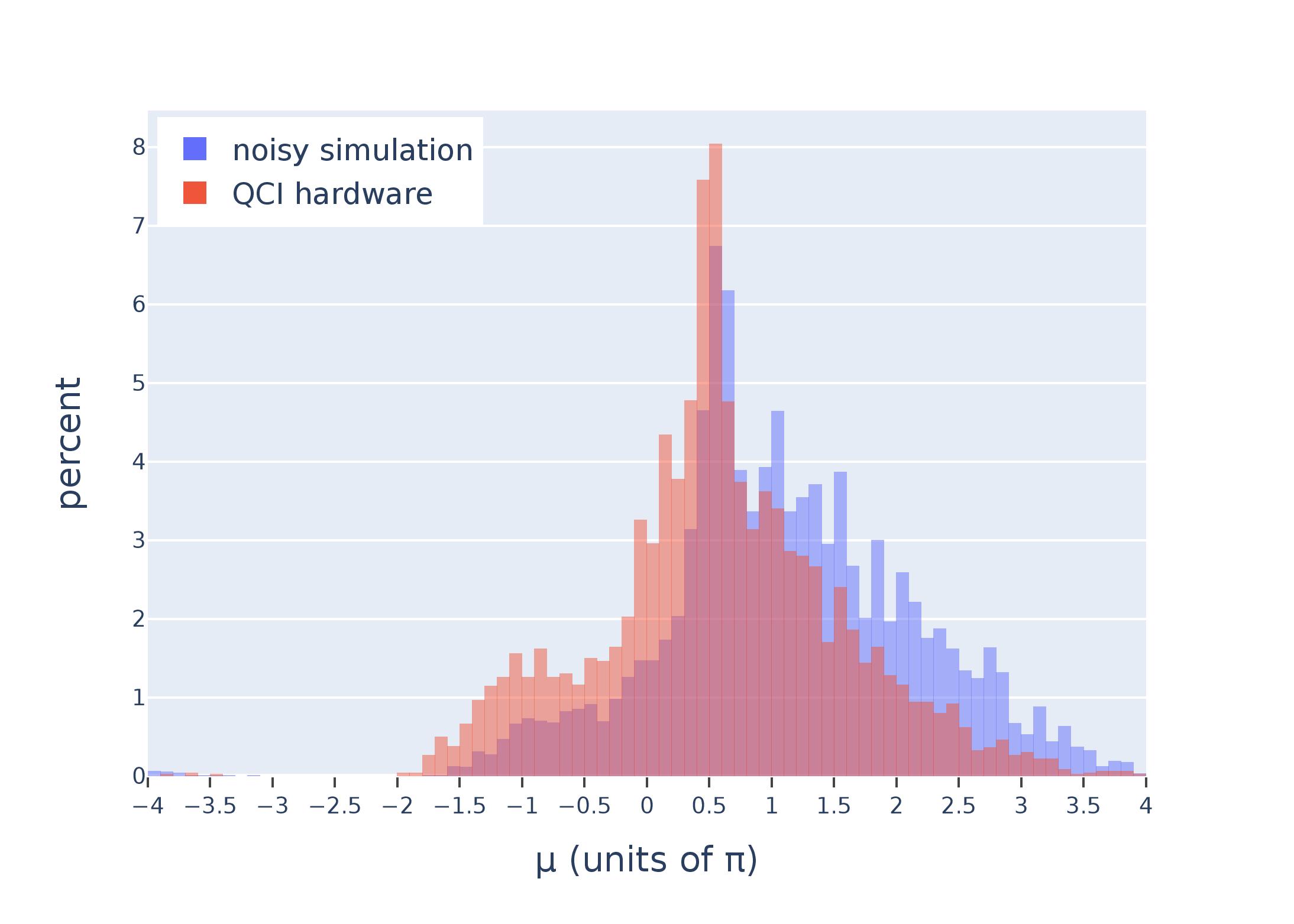}
    \caption{Results on hardware compared with a noisy simulation. We collected 5,000 shots on hardware and 10,000 shots in the simulator. This histogram is comparable with \autoref{fig:ideal_sim} because it uses the same x-axis bounds and number of bins. The correctness of the result is evident in the position on the x-axis of the highest peak, not directly its height. The peak at $\mu=0.5$ is lower in this histogram because the noise made it less likely to converge there.}
    \label{fig:hw_vs_sim}
\end{figure}

In \autoref{fig:hw_vs_sim} we present results from execution of the RWPE program on the QCI quantum hardware alongside results from the quantum simulator described above (\autoref{sec:execution_on_simulator}). 
Execution on the simulator was performed using a QCI-specific noise model along with classical computation which models the fixed-point precision of the system.

For both the noisy simulation and the hardware execution, there is a prominent peak in the data at $\mu=0.5$ which corresponds to the eigenvalue described in \autoref{sec:parameterizing_for_execution}, matching closely what was seen in the ideal simulation.
The correctness of these results is suggested by the location of the highest peak in this histogram.
Both the hardware and simulator results include shoulders in the data, loosely corresponding to where we see population in the ideal simulation in \autoref{fig:ideal_sim}. Detailed error analysis and quantitative comparison of these results was deferred to future work. 
These results may be considered sufficient to confirm a successful translation of the quantum program defined in the QIR representation, resulting in nearly equivalent results on the both the simulator and the QCI hardware. 

\vspace{0.3cm}

The native gate set used in the QCI hardware system includes the \lstinline+H+, \!$\sqrt{\text{\lstinline+X+}}$, \lstinline+X+, and \lstinline+RZ+ single qubit gates. As discussed in \autoref{sec:hardware_challenges}, the \lstinline+RZ+ gate can accept variable arguments at run-time and is executed in a small fraction of the time it takes to execute a pulse-based gate. The native entangling gate in the QCI hardware is the exponential-SWAP~\cite{gao2019entanglement}, or \lstinline+ESWAP+, which can also accept a run-time variable argument, and which has the unitary

\begin{equation}
    \begin{aligned}
        U_{\eSWAP}(\theta) & =
            \begin{pmatrix}
                e^{-i\theta/2} & 0 & 0 & 0 \\
                0 & \cos(\theta/2) & -i\sin(\theta/2) & 0 \\
                0 & -i\sin(\theta/2) & \cos(\theta/2) & 0 \\
                0 & 0 & 0 & e^{-i\theta/2}
            \end{pmatrix} \\
            & = e^{-i\frac{\theta}{2}\left(\openone\openone + XX + YY + ZZ\right)}.
    \end{aligned}
\end{equation}
Useful for larger programs than the one demonstrated here, the \lstinline+SWAP+ gate is a special case of the \lstinline+ESWAP+ corresponding to $\theta=\pi$. Near synonyms of this gate are referred to as the \lstinline+SwapPowGate+~\cite{google_cirq} and ${\rm SWAP}^\alpha$ \cite{fan2005optimal}. Gate fidelity is likely the largest factor influencing the height of the peak in the results for both the noisy simulator and hardware, however, and these fidelities were sufficient for the RWPE algorithm to produce a well-defined solution.


\section{Future Work}
\label{sec:future_work}

The RWPE algorithm presented here could be enhanced with additional study.
We limited our consideration only to those inferences made on timescales significantly shorter than qubit lifetimes. Even under such severe constraints, the deep integration of classical and quantum computation allows us to derive estimates of the phase in real-time. With this unique capability, we can validate hypotheses that mitigate the impact of noise and other errors on phase estimation.

Generally, we could relax this restriction by re-analyzing intermediate measurements after the fact. The sequence $\{(t_i, \phi_{\text{inv},i}, d_i)\}$ of evolution times, inversion angles, and intermediate measurement results is sufficient to capture the inferential effect of the RWPE protocol in the likelihood function:
\begin{align}
    \label{eq:post-proc-like}
    \Pr(\text{data} | \phi) = \prod_i \begin{cases}
        \cos^{2}([\phi - \phi_{\text{inv},i}] t_i / 2) & \text{if } d_i = 0 \\
        \sin^{2}([\phi - \phi_{\text{inv},i}] t_i / 2) & \text{if } d_i = 1.
    \end{cases}
\end{align}
Taking an expectation over a prior distribution updated by (\ref{eq:post-proc-like}) yields an estimate that minimizes the average mean squared error in $\phi$ \cite{banerjeeOptimalityConditionalExpectation2005}; such expectation values can be readily computed in postprocessing using software packages such as PyMC3 \cite{pymc3}~or QInfer \cite{qinfer}. For example, \citet{rwpe_theory} used a QInfer particle filter to post-process results of RWPE execution and observed a reduction in the impact of outliers on overall performance, allowing recovery from some approximation failures observed in \autoref{fig:ideal_sim}.

In a similar fashion, future work could incorporate real-time hybrid applications more deeply into data processing workflows.
Intermediate measurements can be used to inform online experiment design heuristics and approximations, while the full power of classical data processing can be used outside of qubit lifetimes to refine estimates offline.

\vspace{0.3cm}

Beyond the RWPE algorithm, an obvious next step is to explore other quantum algorithms that might benefit from the ability to do mathematical computation within the quantum program.
One use case for this is in the quantum algorithms designed to implement error correction.
Any quantum algorithm that requires some computation of classical variables within the algorithm itself could conceivably be implemented using classical arithmetic operations rather than with quantum gates.
Other known complex algorithms could be candidates for efficiency gains, such as in chemistry simulation \cite{reiherElucidatingReactionMechanisms2017}.
Developing entirely new and novel algorithms that use this capability at their core is another area to be explored.

\vspace{0.3cm}

The section on Code Availability below \ref{sec:data_and_code_availability} provides information about various public repositories containing tools related to the Quantum Intermediate Representation (QIR). We are early in the evolution of this new form of hybrid and adaptive quantum programming and it is likely that additional resources and examples will soon become available.


\section{Summary and Conclusions}
\label{sec:summary-and-conclusions}

In this paper, we have demonstrated an initial step towards fully general and tightly integrated quantum and classical processing, backed by QIR, an intermediate representation that can be used to express hybrid quantum--classical programs in a form that can be translated to the unique assembly language of specific hardware targets.
We then used a newly developed suite of supporting software tools together with advanced quantum simulation and physical quantum hardware capabilities to demonstrate random walk phase estimation, a recent algorithm that effectively exploits real--time hybrid quantum--classical computation within a quantum program.

The Quantum Algorithm Zoo \cite{jordan_zoo} lists 64 quantum computing algorithms --- by comparison, Knuth's unfinished multi-volume compendium of classical algorithms \cite{knuth_2011} has more than 64 chapters. One hypothesis for the enormous gap between the number of quantum and classical algorithms is that we have a better language for the latter, as it is easier to conceive of new algorithms when we have a language in which to express them \cite{ive_notation_1980} easily, along with a robust software and hardware ecosystem in which they can be exercised.

If it is the case that progress towards development of new and more efficient quantum algorithms has been limited by our computational models, then the work described in this paper offers substantial progress towards resolving that gap.
We hope that the results that we demonstrated in this paper provide an impetus to fundamentally rethink and expand what a quantum algorithm can look like and to go beyond the limitations seen in the current algorithms.

The success of quantum computing hinges not only on progress in qubit fidelities and lifetimes and in electronics for controlling qubits, but also on the development of innovative software components that support the hardware in creative and extensible ways.
Our work may enable us to narrow the gap, and accelerate the development of new techniques.


\section*{Code Availability}
\label{sec:data_and_code_availability}

The QIR Alliance \cite{qir_alliance} provides information and tools to support the development of a specification for Quantum Intermediate Representation (QIR) \cite{qir_spec}.
Source code for tools such as ``PyQIR'' and ``QAT'' mentioned in the text of this paper is available online \cite{pyqir_2022,qat_2022}.

For more information about the OpenQASM 3.0 specification, please see \citet{cross2021openqasm}.

\clearpage
\appendix
\clearpage

\section{Implementation Detail and Analytics}
\label{apdx:algorithms_and_applications}


\subsection{Source Code}
\label{apdx:source-code}

The Q\# source code for the RWPE advanced hybrid quantum algorithm is shown below.
The code largely matches the description in the text of this paper with a few exceptions.

The main entry point for the RWPE algorithm is shown in \autoref{lst:program-qs}.
Intrinsic function definitions for the CRz and ActiveReset operations are shown in \autoref{lst:custom-qs}.
The \qsharp~project file is shown in \autoref{lst:project-qs}.

QCI's hardware has a multi-qubit reset protocol, modeled in \autoref{alg:qubit_reset}, which is more effective than multiple single qubit reset operations and which we wanted to use in this experiment. Instead of having our QIR translator look for a sequence of multiple single qubit resets, we added an \lstinline+ActiveReset+ instruction which may be invoked from the \qsharp~code using the definition in \autoref{lst:custom-qs}. This \lstinline+ActiveReset+ is only currently usable when all qubits are included, and as such, is not appropriate for the single qubit reset in the circuit. 

\begin{figure*}
    \includegraphics[width=0.9\textwidth]{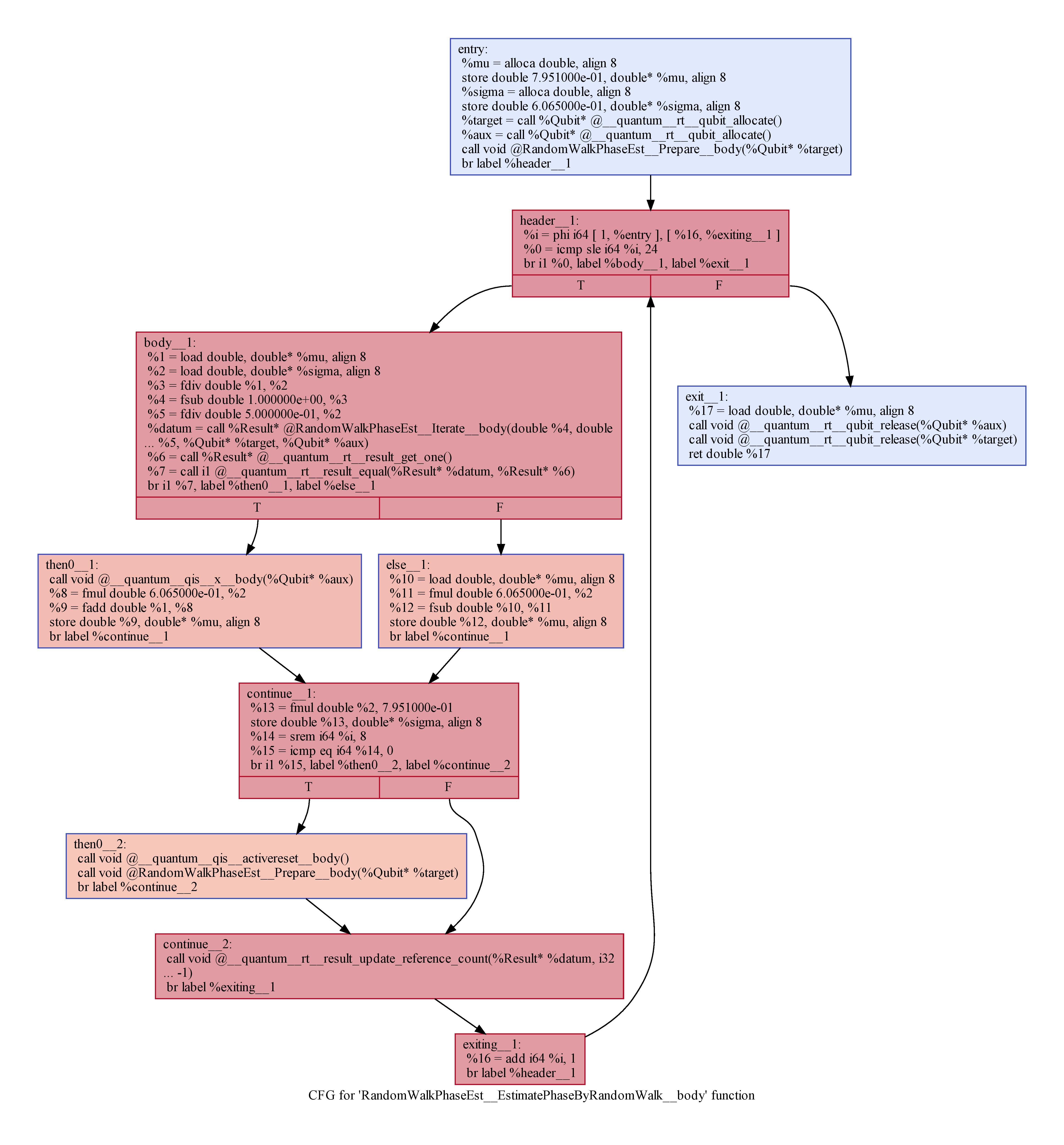}
    \caption{
        Control flow graph for compiled from the random walk phase estimation implementation in \autoref{apdx:source-code}. Note that in the \lstinline+body__1+ block, \lstinline+\%1+ and \lstinline+\%2+ represent the current values of $\mu$ and $\sigma$ within each iteration of the main phase estimation loop, and are used to compute the inversion angle and evolution time passed to the phase estimation iteration function. The two branches of the random walk are then represented by \lstinline+then0__1+ and \lstinline+else__1+.
    }
    \label{fig:rwpe-cfg}
\end{figure*}

When this Q\# program is compiled into QIR, the result is a body of LLVM code that describes the program, but in the intermediate form that can readily be processed by disparate hardware providers for execution on specific backend systems.
The LLVM code generated for the RWPE program is shown in below \autoref{fig:rwpe-cfg}.

\onecolumngrid

\vspace{0.3cm}

\begin{lstlisting}[
    language=qsharp,
    caption={\texttt{Program.qs}: Main entry point for \qsharp~implementation of \autoref{alg:rwpe-no-unwinding}.},
    label={lst:program-qs}
]
namespace RandomWalkPhaseEst {
    open Microsoft.Quantum.Intrinsic;
    open Microsoft.Quantum.Math;
    open Microsoft.Quantum.Measurement;

    operation Prepare(target : Qubit) : Unit {
        X(target);
    }

    operation Iterate(c1 : Double, c2 : Double, target : Qubit, aux : Qubit) : Result {
        within {
            H(aux);
        } apply {
            Rz(c1, aux);
            // NB: This implicitly defines the oracle $U$ provided as input
            //     in Algorithm 3 to be $R_z$.
            CRz(c2, aux, target);
        }
        return M(aux);
    }

    @EntryPoint()
    operation EstimatePhaseByRandomWalk() : Double {
        let nIter = 24;
        mutable mu = 0.7951;
        mutable sigma = 0.6065;

        use target = Qubit();
        use aux = Qubit();

        Prepare(target);

        for i in 1 .. nIter {
            // NB: The 0.5 on this line sets the scale of the oracle $U$, such
            //     that $\phi = 0.5$ is the true eigenvalue that we wish to
            //     find.
            let datum = Iterate(1.0 - (mu / sigma), 0.5 / sigma, target, aux);

            if datum == One {
                X(aux);
                set mu += 0.6065 * sigma;
            }
            else {
                set mu -= 0.6065 * sigma;
            }
            set sigma *= 0.7951;

            if i % 8 == 0 {
                ActiveReset();
                Prepare(target);
            }
        }

        return mu;
    }
}

\end{lstlisting}

\begin{lstlisting}[
    language=qsharp,
    caption={\texttt{Custom.qs}: Custom instructions used in \autoref{lst:program-qs}.},
    label={lst:custom-qs}
]
// Copyright (c) Microsoft Corporation. All rights reserved.
// Licensed under the MIT License.

namespace Microsoft.Quantum.Intrinsic {
    open Microsoft.Quantum.Targeting;
    open Microsoft.Quantum.Math;

    @TargetInstruction("crz__body")
    operation CRz (theta : Double, control : Qubit, target : Qubit) : Unit {
        body intrinsic;
    }

    operation ActiveReset() : Unit {
        body intrinsic;
    }

}
\end{lstlisting}

\begin{lstlisting}[
    language=xml,
    caption={\texttt{RandomWalkPhaseEst.csproj}: Project file for \qsharp~implementation of \autoref{alg:rwpe-no-unwinding}.},
    label={lst:project-qs}
]
<Project Sdk="Microsoft.Quantum.Sdk/0.20.2111176282-alpha">

  <PropertyGroup>
    <OutputType>Exe</OutputType>
    <TargetFramework>netcoreapp3.1</TargetFramework>
    <ExecutionTarget>qci.machine1</ExecutionTarget>
  </PropertyGroup>

</Project>
\end{lstlisting}

\vspace{0.3cm}

\vspace{0.3cm}


\clearpage
\twocolumngrid

\bibliographystyle{unsrtnat}  
\bibliography{references}

\end{document}